\documentclass[aps,superscriptaddress]{revtex4}

\usepackage{graphicx}
\usepackage{float}
\usepackage{bm}
\usepackage{hyperref}
\usepackage{amsmath}
\usepackage{txfonts}
\usepackage{epstopdf}
\usepackage{graphicx,pstricks}
\usepackage{amsmath,amssymb}
\usepackage{amsfonts}
\newcommand{\hypgeo}[2]{%
  {\vphantom{F}}_{#1}\kern-\scriptspace F_{#2}%
}
\usepackage{natbib}
\begin{document}
\title{Pair Production  in time-dependent Electric field at Finite times 
\footnote{Abstract of presentation on $30^{th}$ September 2022 in QED Laser Plasmas International Workshop $26 - 30$ September 2022  at Max Planck Institute for the Physics of Complex Systems, Dresden,Germany}}
\author{Deepak\footnote{Corresponding author.\\E-mail address: deepakk@rrcat.gov.in (Deepak).}}
\author{Manoranjan P. Singh}
\affiliation{Theory and Simulations Lab, Raja Ramanna Centre for Advanced
Technology,  Indore-452013, India}
\affiliation{Homi Bhaba National Institute, Training School Complex, Anushakti Nagar, Mumbai 400094, India}
\maketitle
\section{Introduction}
A strong classical electromagnetic background can lead to vacuum instability and produce particle-antiparticle pairs. This process of particle creation from the quantum vacuum was first studied in 1951 by Schwinger under a constant electric field, and this phenomenon is known as the Schwinger effect \cite{PhysRev.82.664}.This particle creation paradigm has crucial importance for nonequilibrium processes in heavy-ion collisions\cite{24,25,26} as well as astrophysical phenomena\cite{27}and the search for nonlinear and nonperturbative effects in ultraintense laser systems\cite{12,13,14}.
\newline
Particle production is the process of evolving a quantum system from an initial equilibrium configuration to a new final equilibrium configuration via an intermediate  non-equilibrium evolution caused by a strong field background.
Such strong field QED pair production process has been investigated using scattering calculations \cite{Harvey:2009ry}, exact solutions \cite{Dunne:1998ni}, semi-classical techniques \cite{Kim:2000un,DiPiazza:2004lsj}, Monte Carlo simulations \cite{Gies:2005bz} and quantum kinetic equations \cite{Alkofer:2001ik,Blaschke:2005hs}
Quantitative description of particle production at all times in time-dependent electromagnetic field is not possible due to  the absence of unique separation into positive and negative energy states at intermediate times and these positive and negative states   are well-defined  only at asymptotically early and late times where  the field vanishes. A common approach is to define particle numbers in terms of an adiabatic basis using Bogoliubov transformation \cite{2,3,4,5}. 
In the adiabatic basis, we examine the problem of pair production in a time-varying spatially uniform Sauter electric field  which has been studied by various authors \cite{6,44}. who calculated  the  number of particles created  at  the asymptotic time but the problem of particle  production at the finite time is not studied for Sauter-Pulse electric field.
We looking for the evolution of the quantum system at some initial time $t_0$ in the vacuum state but now what will be the properties of the quantum system at finite time $t$ ?  What happens to the system properties at that non-asymptotic time? 
By finding the exact analytic solution of the mode function, which allows us to study the behaviour of the quantum vacuum during pair production at any instant of time for time-dependent Sauter pulsed electric fields in spinor quantum electrodynamics (QED).
We use  natural units and set $ \hslash = c = m = 1 $ and express all variables in terms of the electron mass unit.

\section{Theory}
Consider the creation of the electron-positron pair from vacuum  by a linearly polarized  time-dependent spatially uniform  electric field along the $z$-axis which is characterized by the four-vector potential $A_\mu(t) = (0, 0, 0, A(t))$ with $E(t) = - \dot{A} (t).$ 
In our calculations,  we choose the Sauter field in the direction of $z-$axis, which is given by
\begin{equation}
    E(t) = - \dot{A} (t)= E_0 sech^2(t/\tau)
\end{equation}
and corresponding vector-potential
\begin{equation}
    A(t) = -E_0  \tau tanh(t/\tau).
\end{equation}
where $E_0$ amplitude of the electric field and $\tau$ pulse duration of the electric field.
The one-particle momentum distribution function $f(\textbf{p},t)$ is an important quantity in the description of the particle production process in the time-dependent electric field.
As it was demonstrated in \cite{41,42}, f(\textbf{p},t) can be written in terms of the mode solution as
\begin{equation}
 f(\textbf{p},t) = \frac{{ (\omega + \Pi) |\dot\Psi^{(\pm)} \pm i \omega \Psi^{(\pm)}|^2}}{2 \omega}
 \label{f}
\end{equation}
where the mode equation \cite{40,43} 
\begin{equation}\label{dir_eq1}
\ddot\Psi^{(\pm)}+[\omega^2(t)\pm ie\dot A(t)]\Psi^{(\pm)}=0
\end{equation}
with 
\begin{equation}\label{omega}
\omega(t)^2=\Pi^2+\epsilon_\perp ^2,\quad \Pi=p_z+eA(t),\quad \epsilon_\perp^2=\textbf{p}_\perp^2+m^2.
\end{equation}
For Sauter-Field, we are able to exactly solve the mode equation \ref{dir_eq1}, which becomes the hyper-geometric differential equation and its solution is given in terms of hypergeometric functions. Using this solution, we calculate the one-particle distribution function in terms of hypergeometric functions. The time evolution of the particle distribution function $f(\textbf{p},t)$ in momentum space is studied for $E_0 = 0.2  $ and $\tau = 10.$
\begin{figure}[t]
\begin{center}
{
\includegraphics[width = 2.5in]{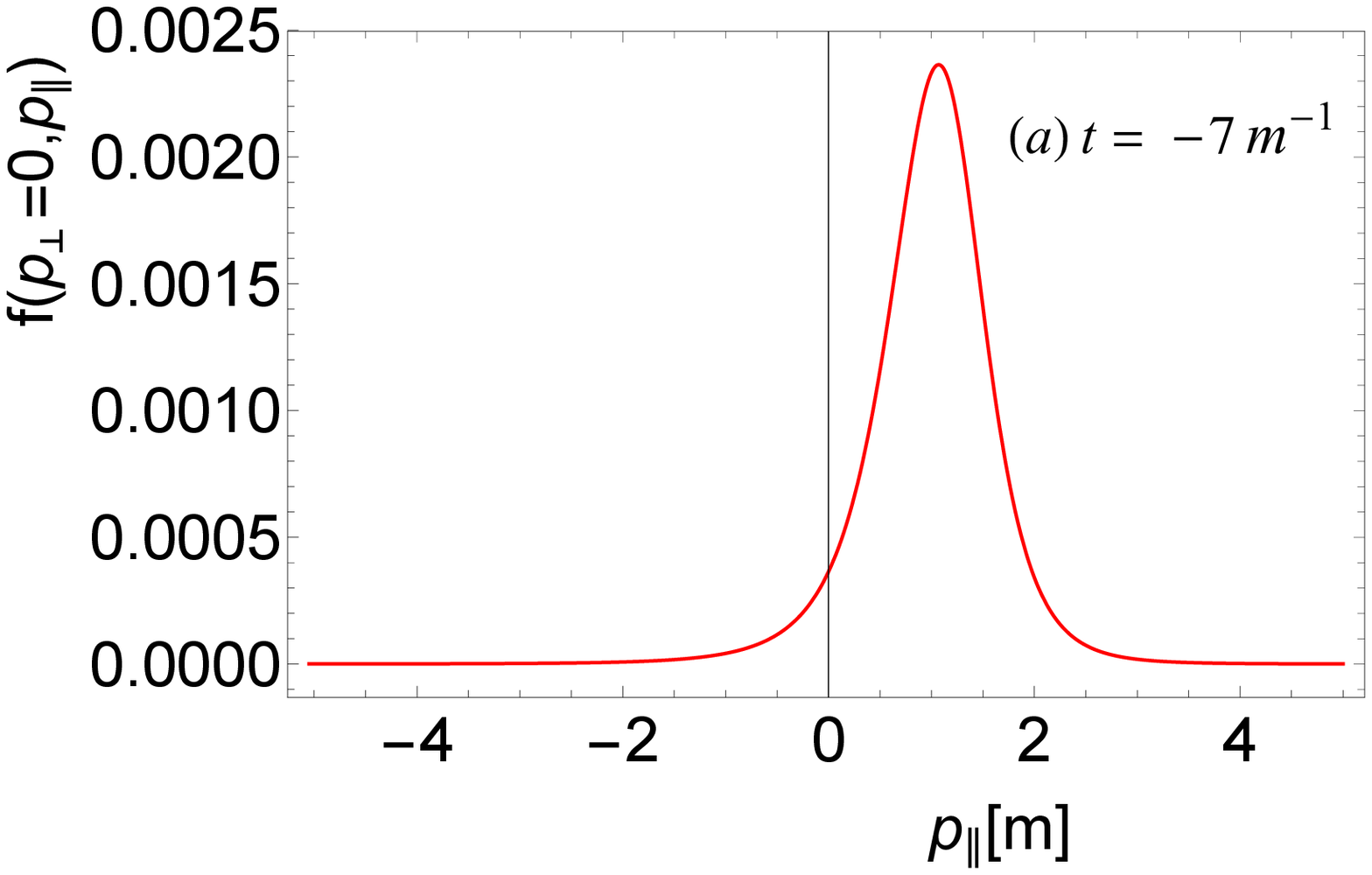}
\includegraphics[width = 2.5in]{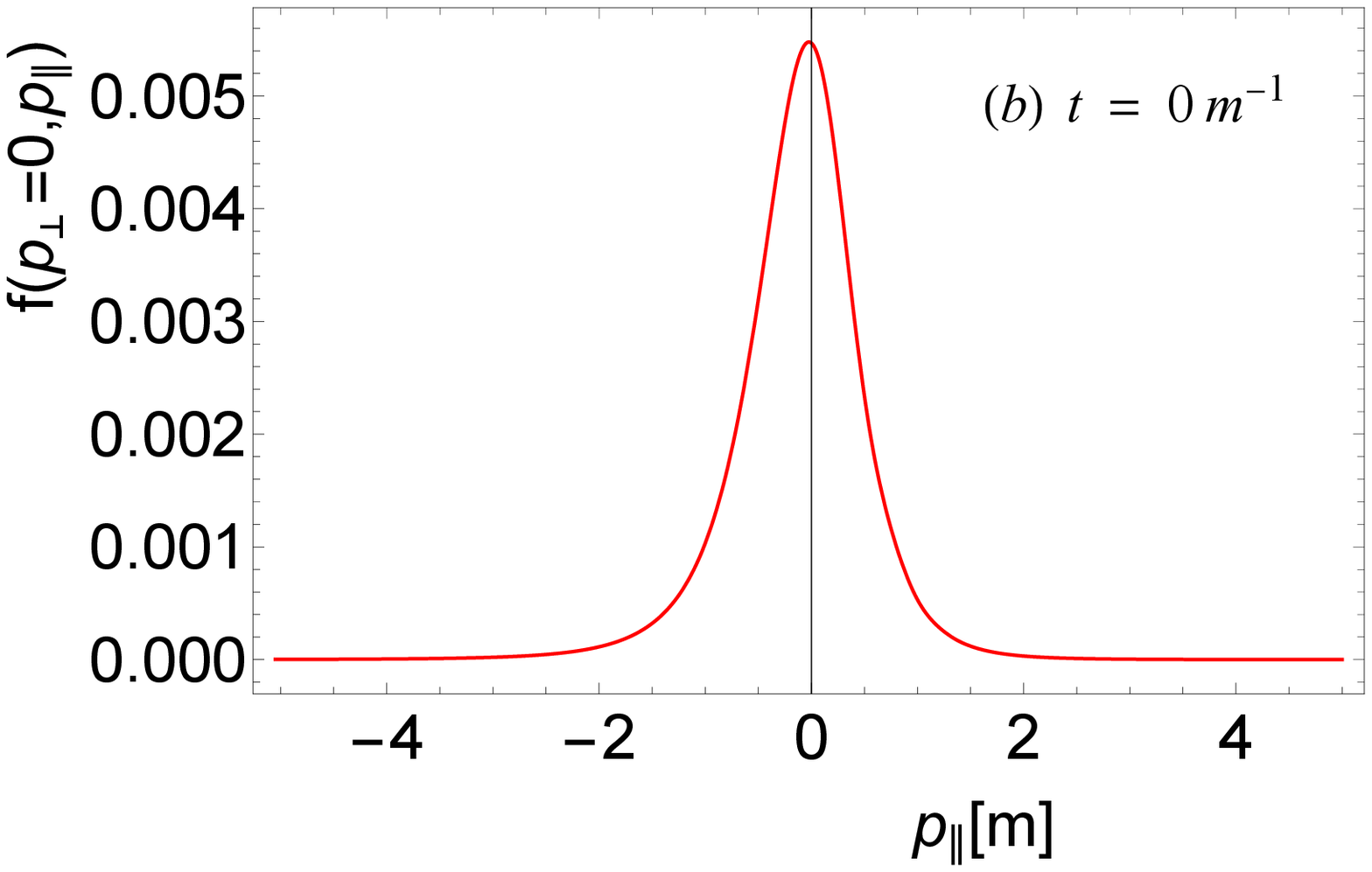}
\includegraphics[width = 2.5in]{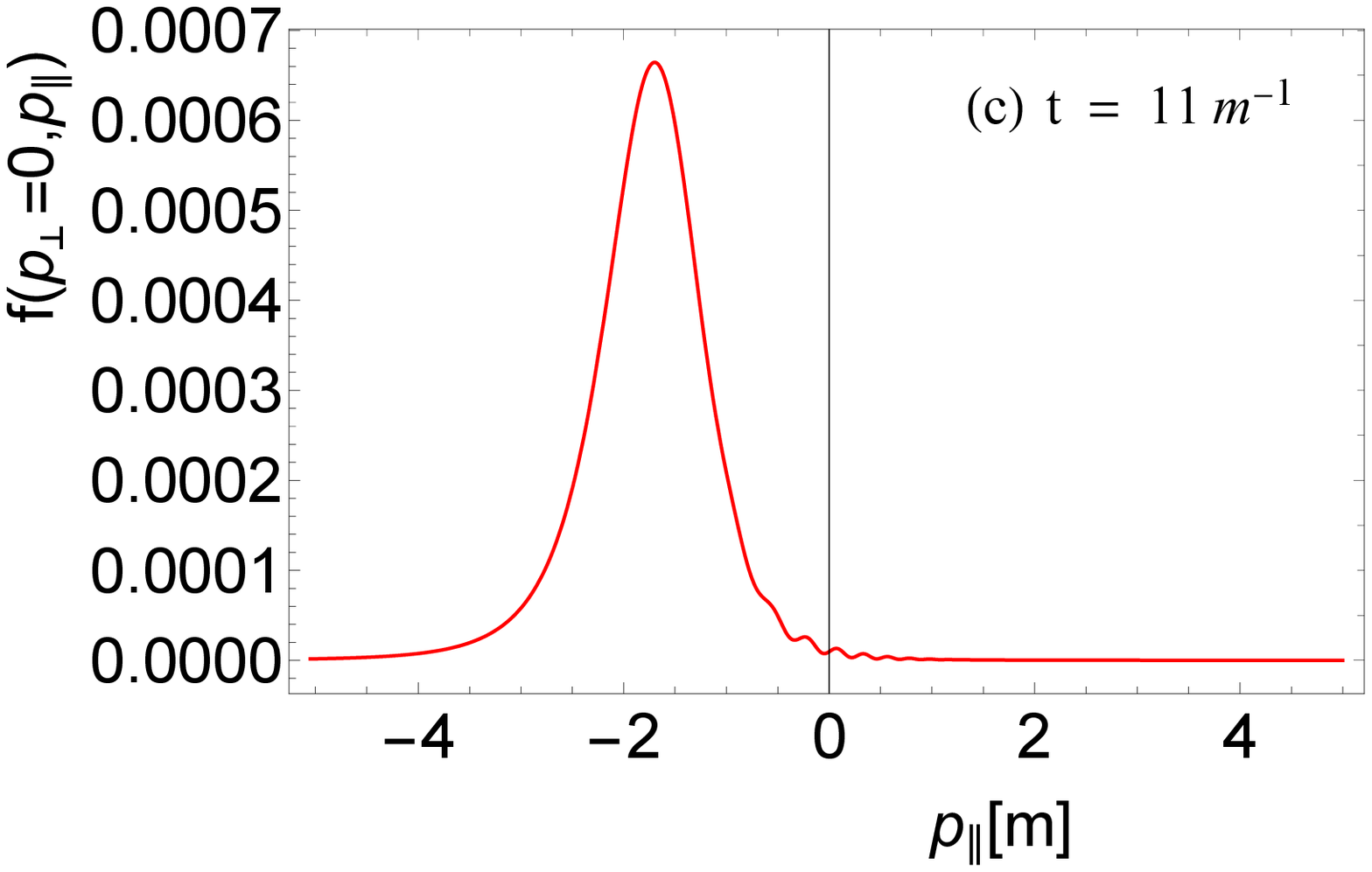}
\includegraphics[width = 2.5in]{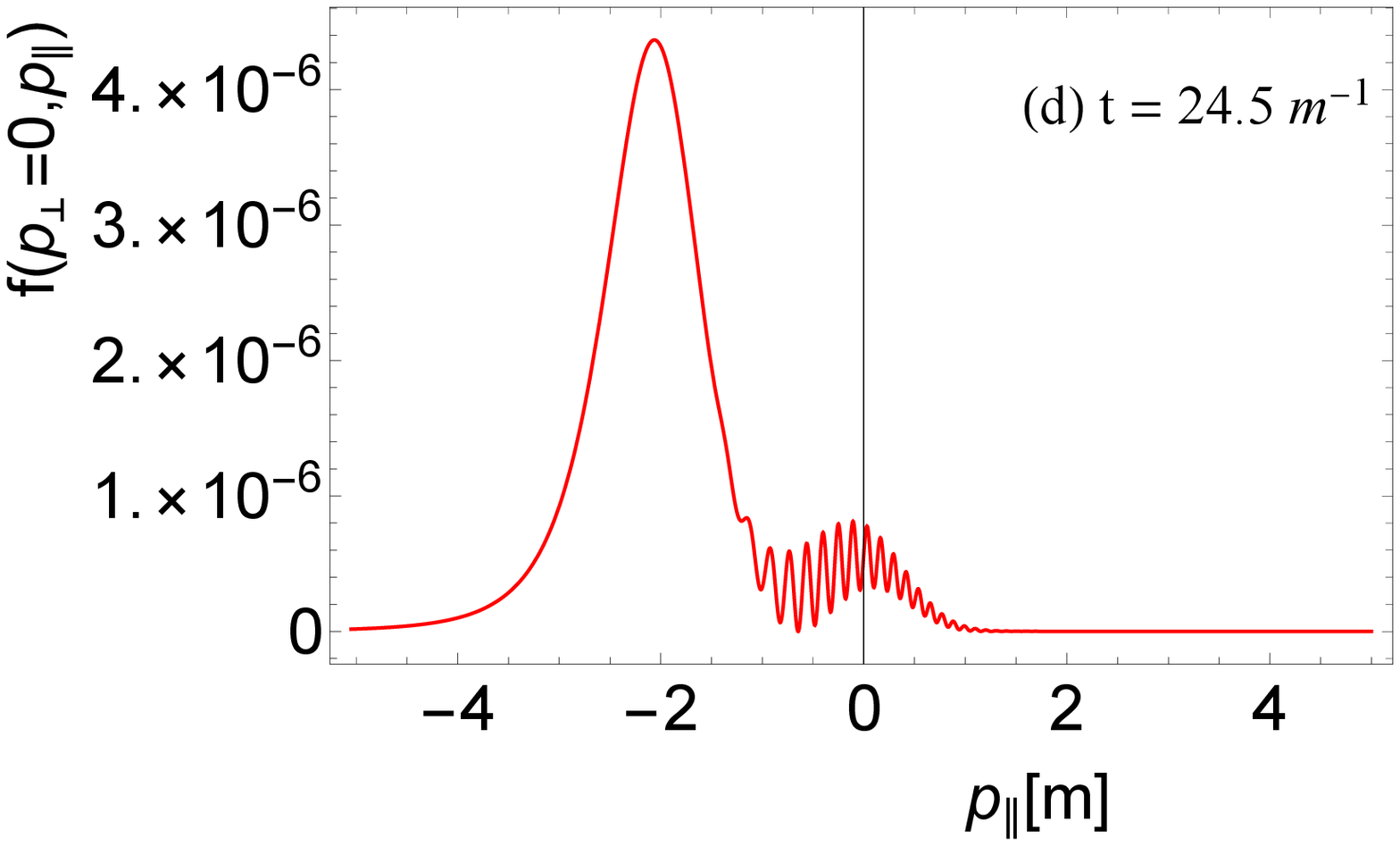}
\includegraphics[width = 2.5in]{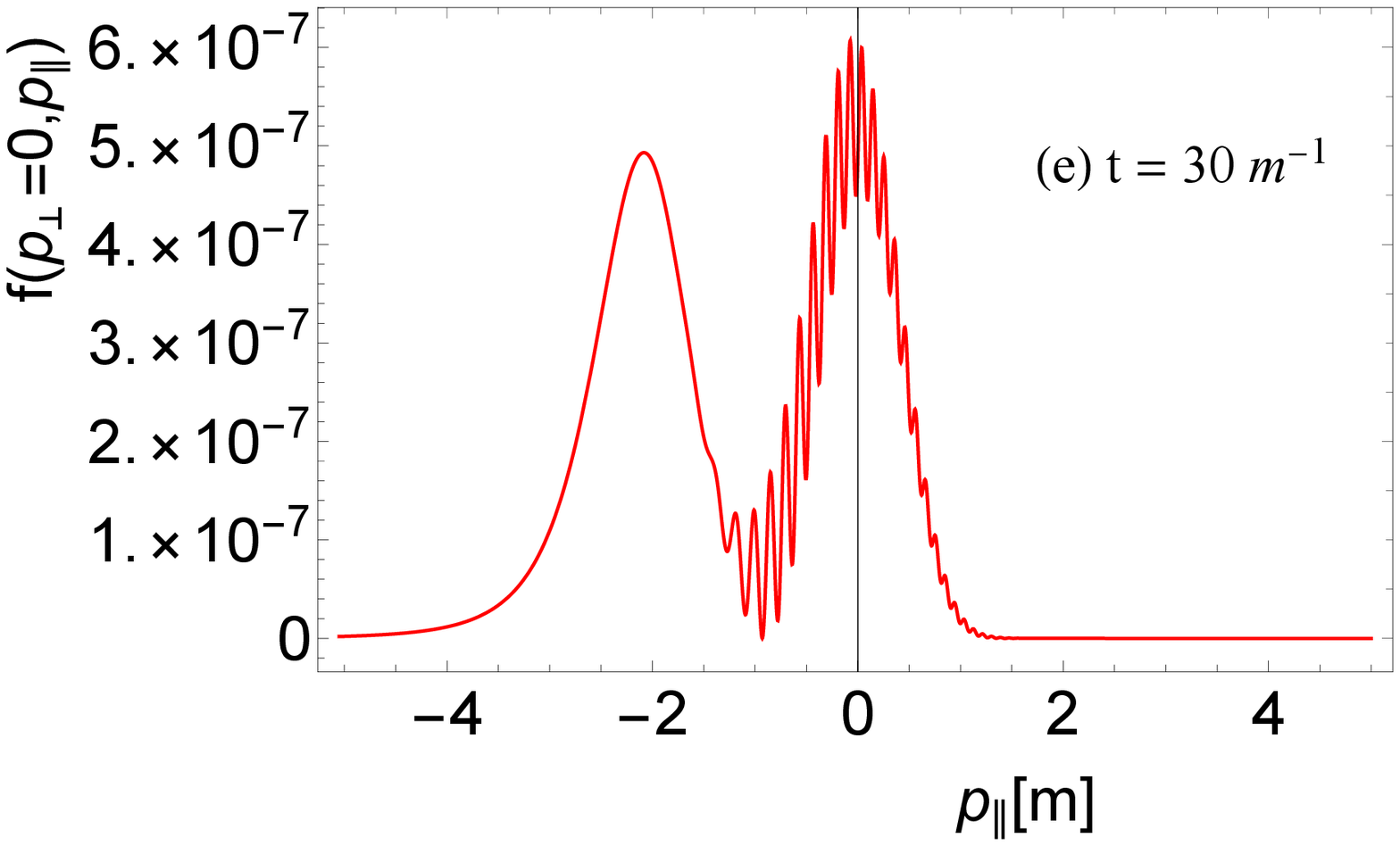}
\includegraphics[width = 2.5in]{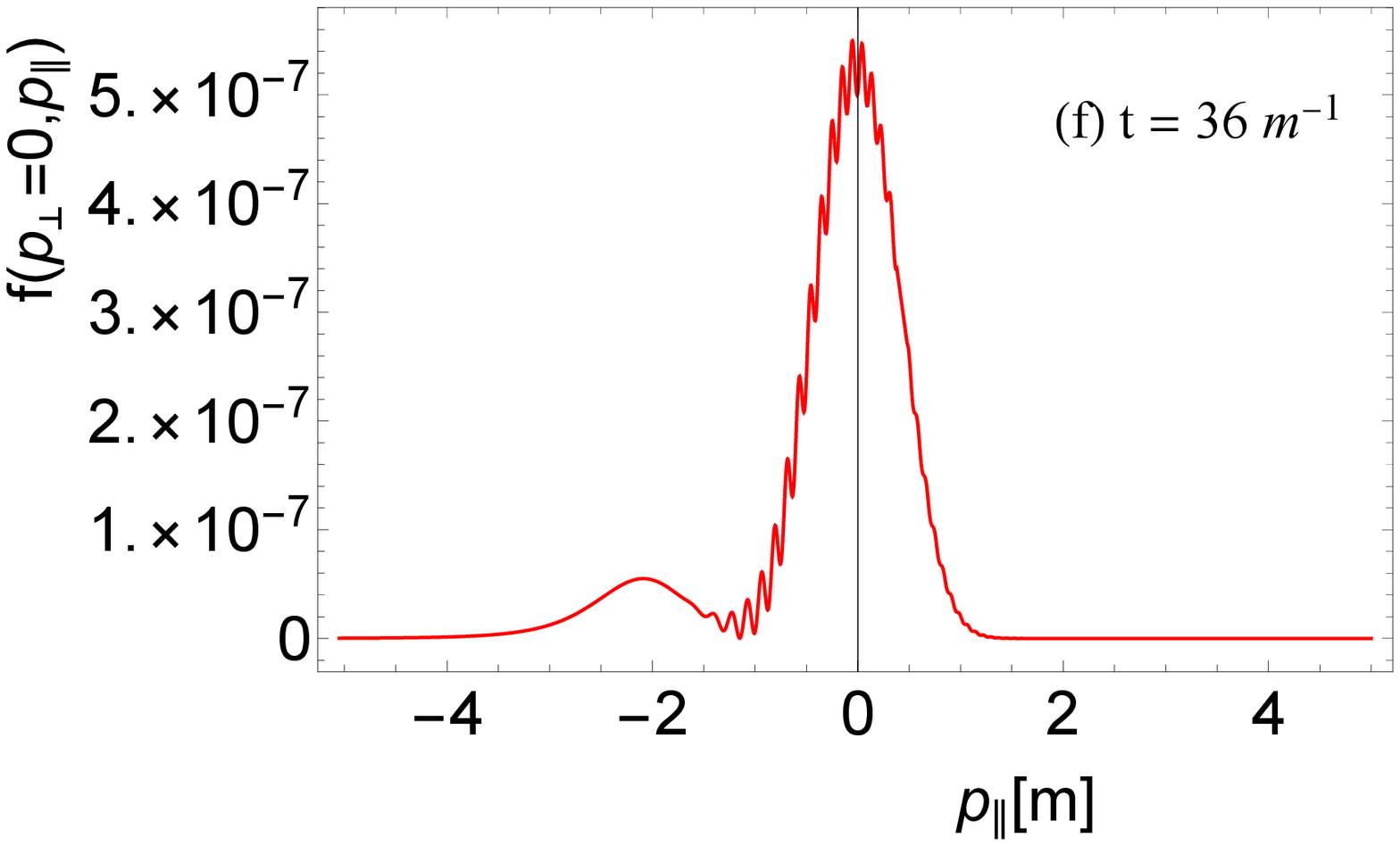}
\includegraphics[width = 2.5in]{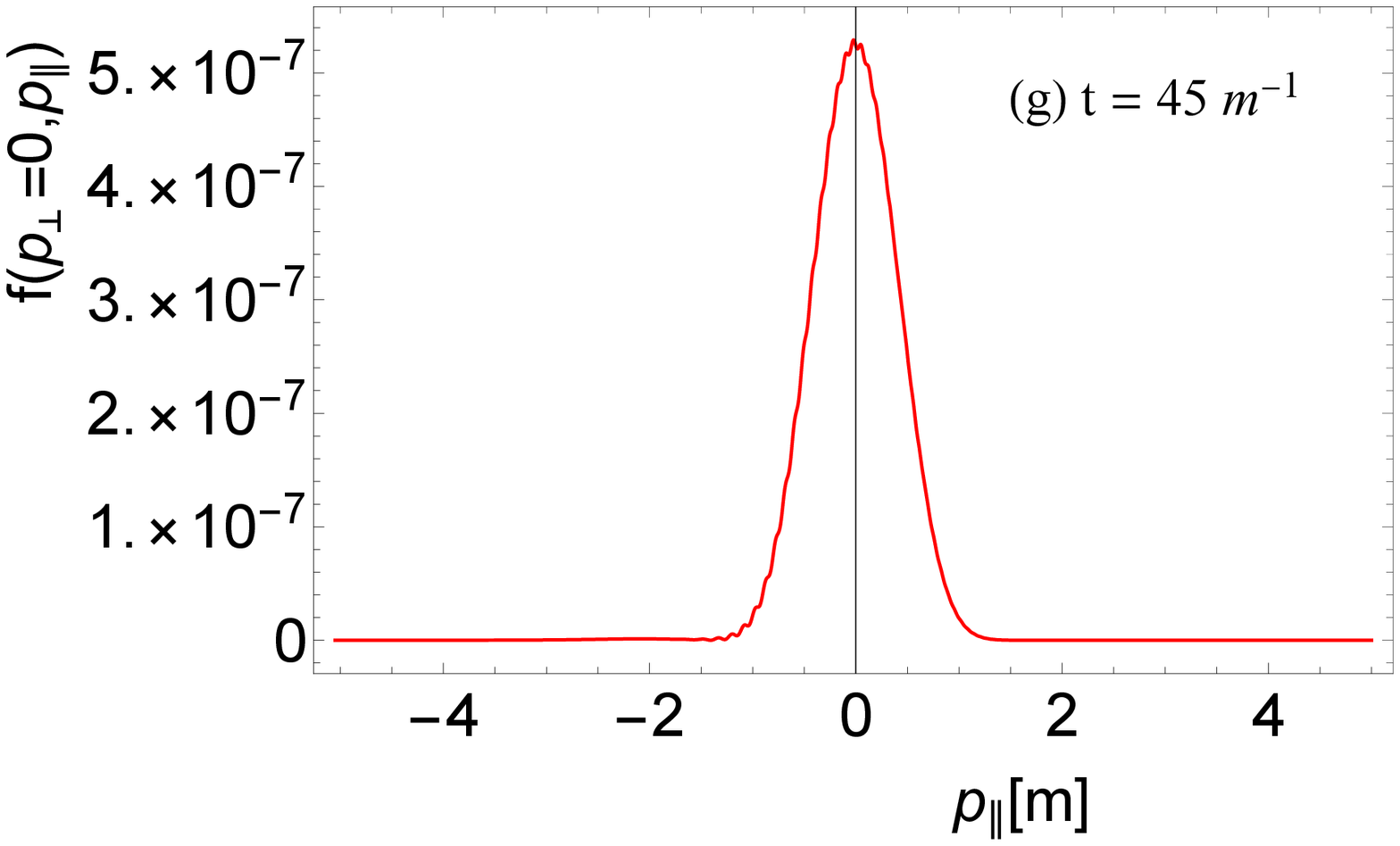}
\includegraphics[width = 2.5in]{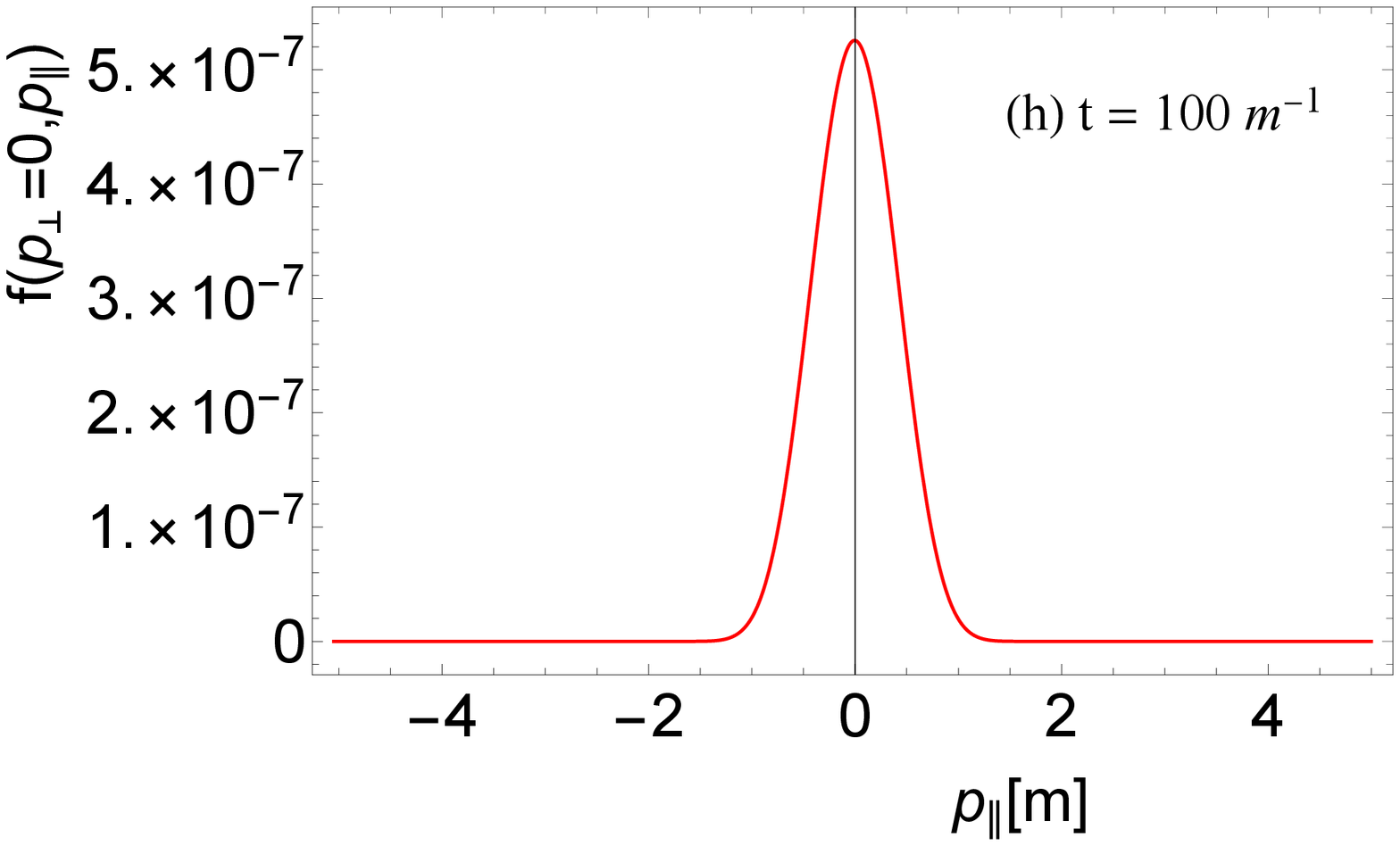}
}
\caption{Time evolution of quasi-particle distribution function in   the longitudinal momentum space $f(p_\parallel, t)$  at different times. The transverse momentum is considered to zero, and all the units are taken in the electron mass units.The field parameters are  $E_0=0.2 E_c$ and $ \tau =10.$}
   	\label{fig:1}
\end{center}
\end{figure}
\section{Result}\label{Result}
In this section, we discuss about both the longitudinal and transverse  momentum spectrum of the created particle using the solution of the mode equation in \ref{f} for $E_0 = 0.2$ and $ \tau = 10 $.

\subsection{Longitudinal momentum spectrum}
In Figure \ref{fig:1}, we show how the particle-creation proceeds. The produced electron is accelerated to the negative z-direction by the electric field. At early times (QEPP region \cite{46}) when electric field$E(t)$ is increasing, momentum distribution shows the smooth Gaussian-like structure and after reaching  a maximum value at time $t=0$ electric field  starts decaying,  and we see the peak of the momentum distribution function  is shifted as shown in figure \ref{fig:1}(a),(b),(c).
As time progresses, the distribution function tends to be peaked around the zero value of the longitudinal momentum as seen in figure \ref{fig:1}(d). In this process, it shows a complex behavior of splitting and manifests oscillation arising at $t = 3 \tau $ (near REPP region ) where the electric field nearly vanished see figure \ref{fig:1}(e),(f). These complex behavior disappear at the asymptotic time ( $t= 10 \tau$), and  we see a smooth Gaussian-like structure \ref{fig:1}(h).
\begin{figure}[t]
\begin{center}
{
\includegraphics[width = 2.4755in]{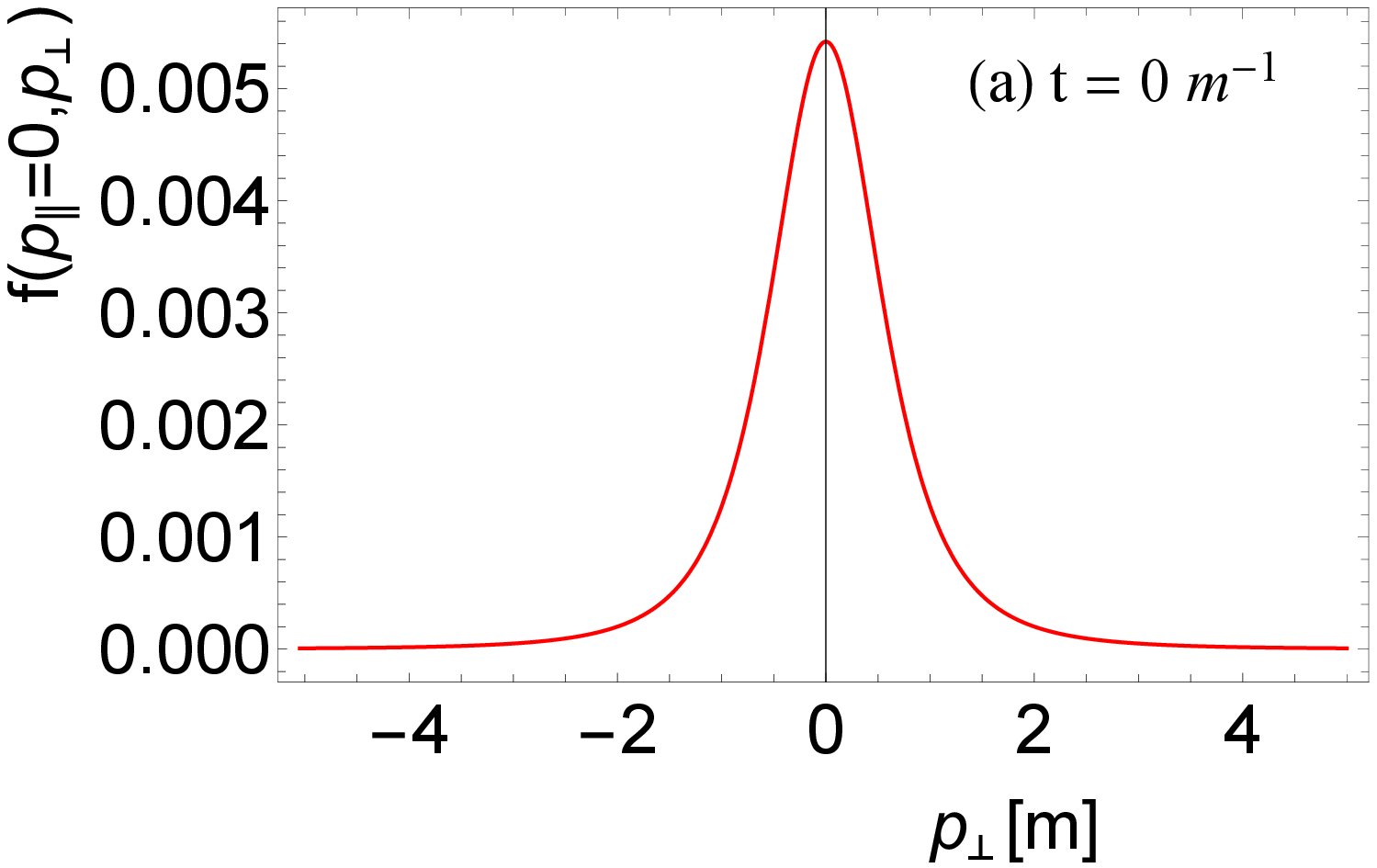}
\includegraphics[width = 2.55in]{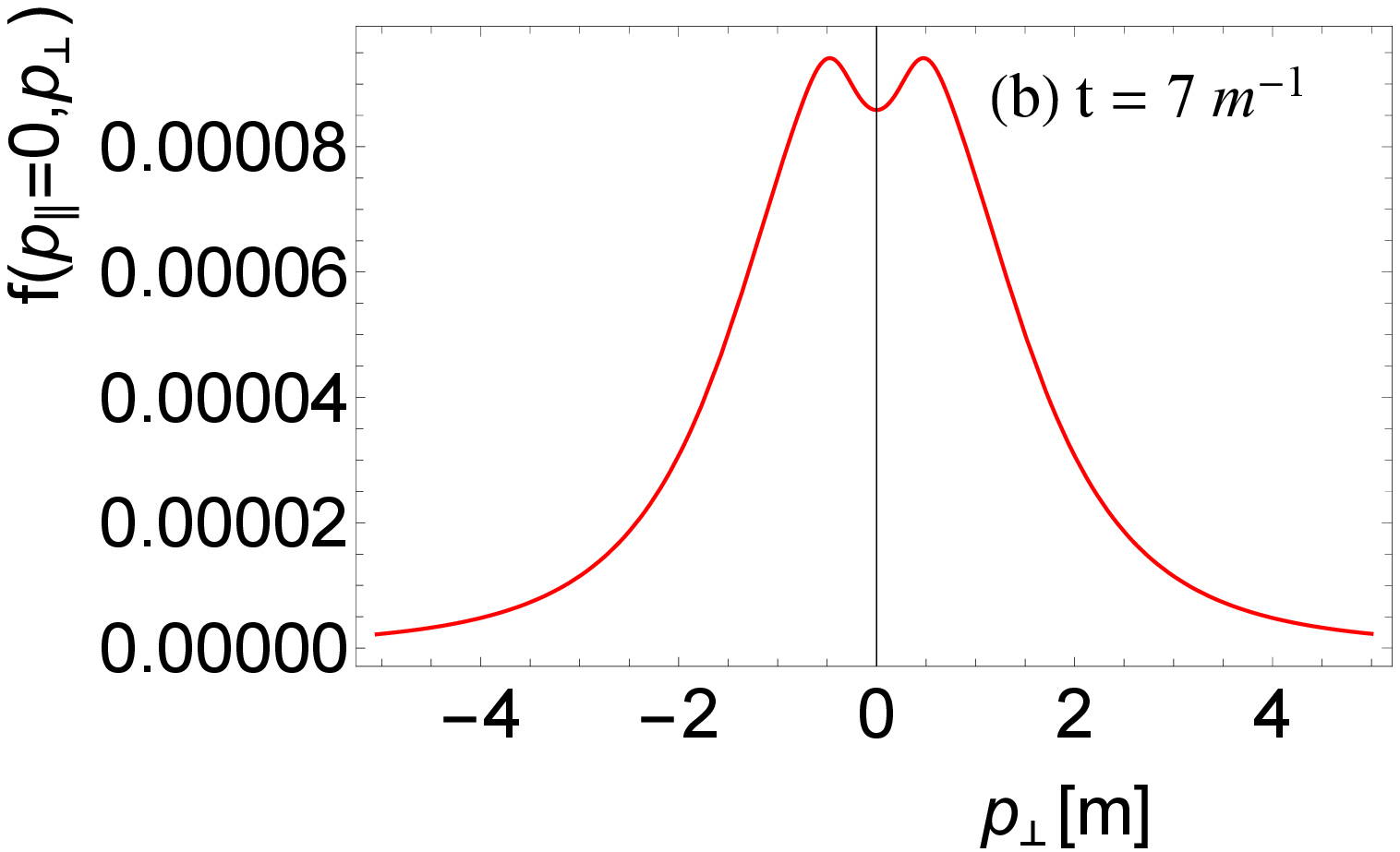}
\includegraphics[width = 2.55in]{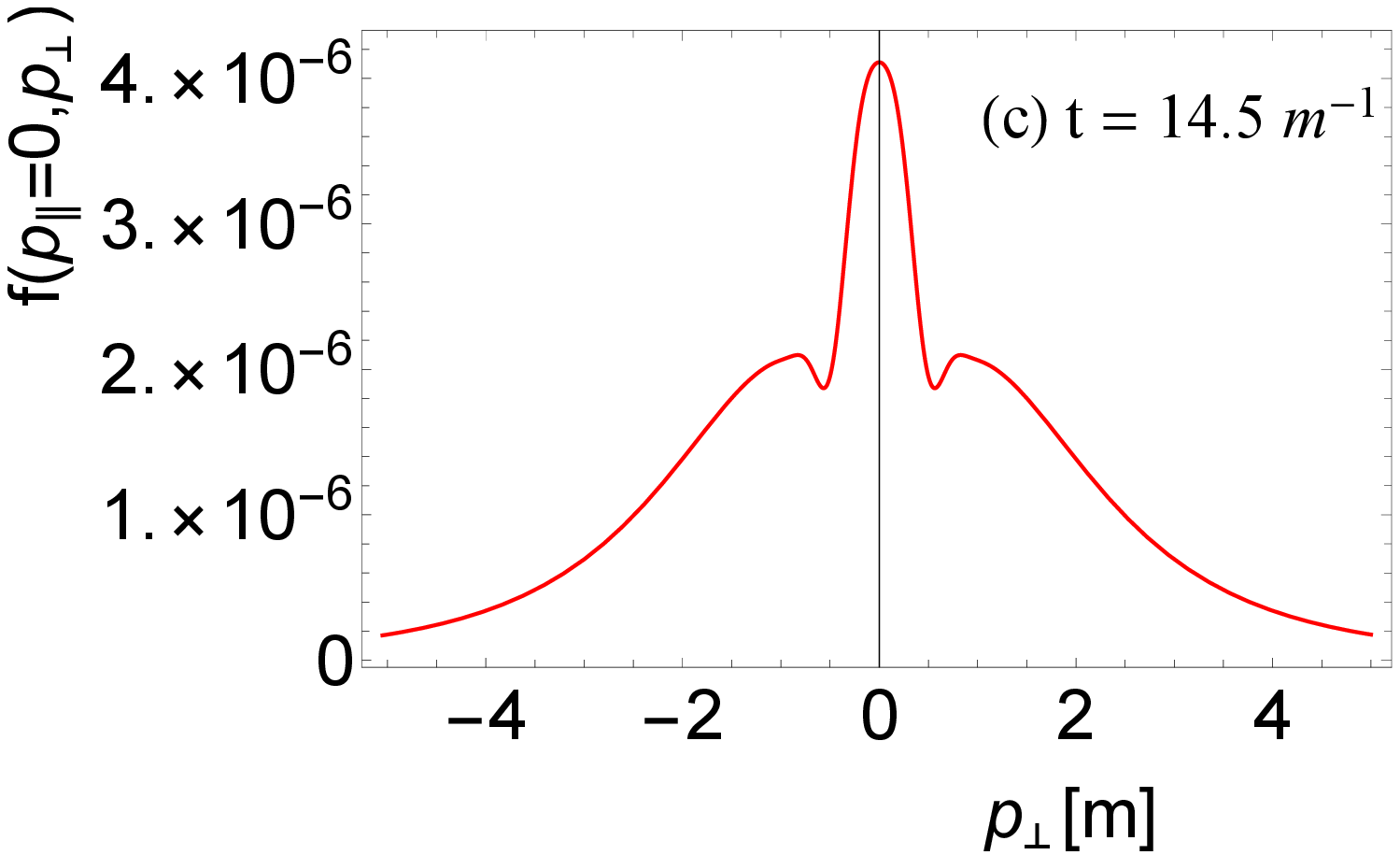}
\includegraphics[width = 2.55in]{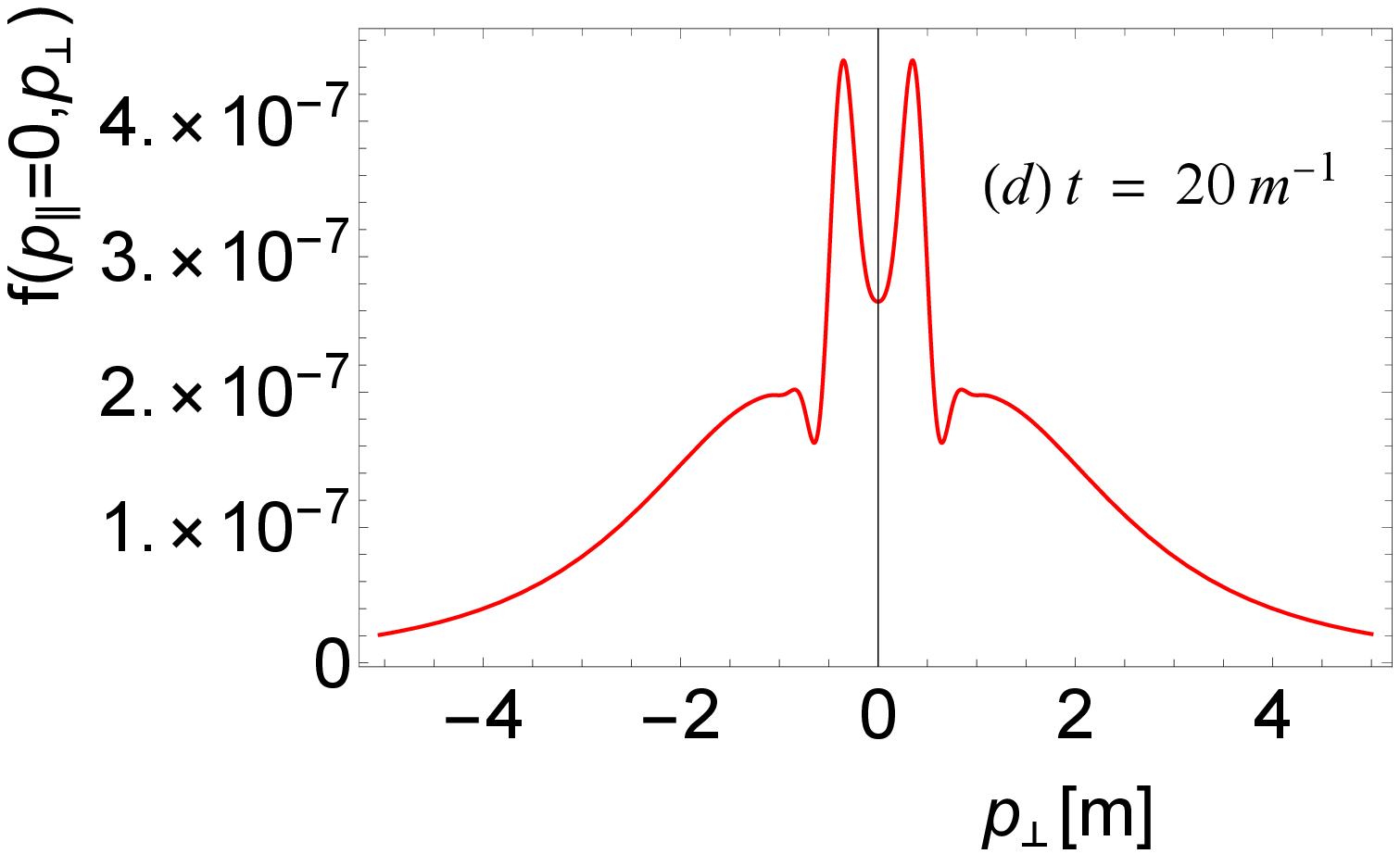}
\includegraphics[width = 2.55in]{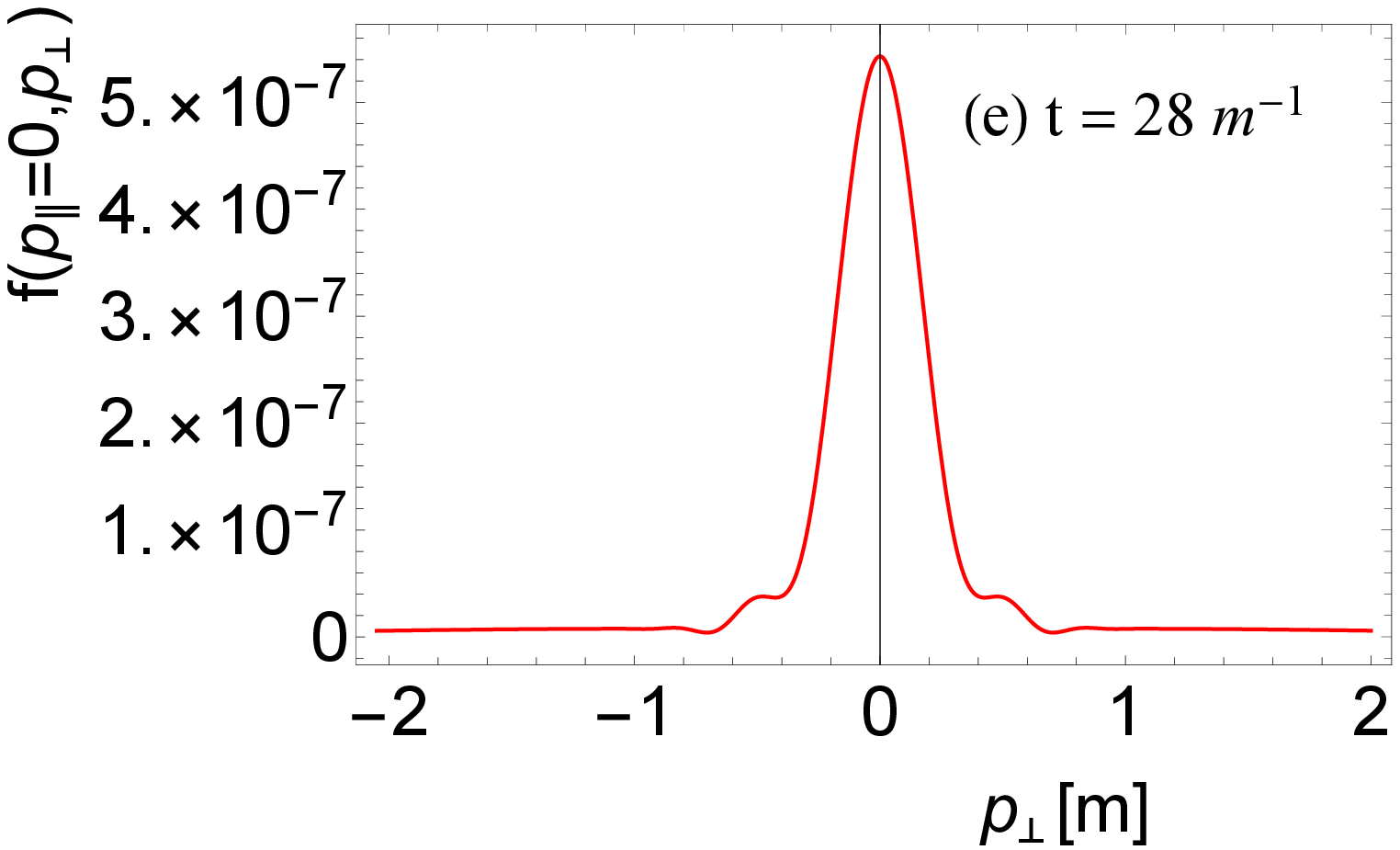}
\includegraphics[width = 2.55in]{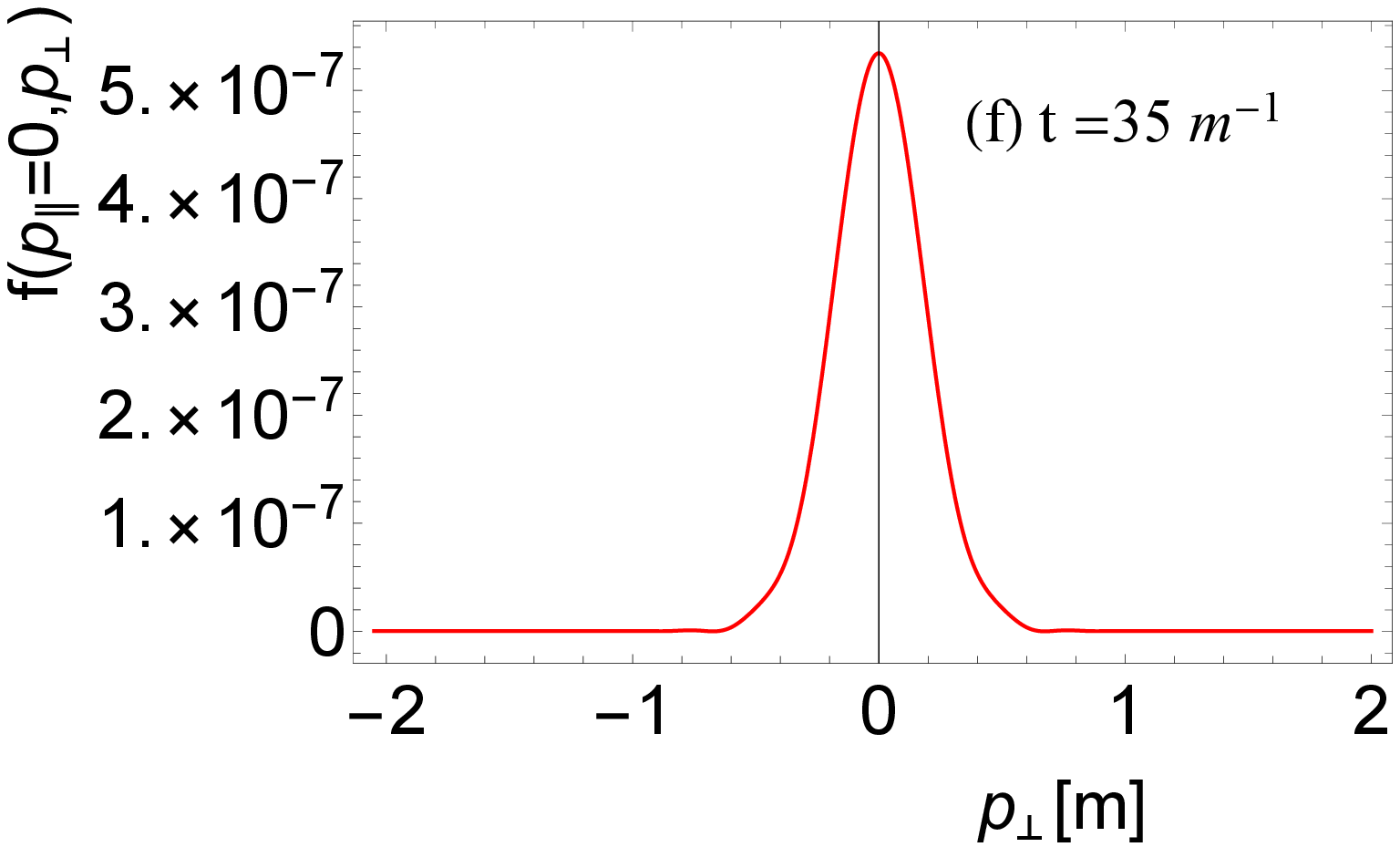}
\includegraphics[width = 2.55in]{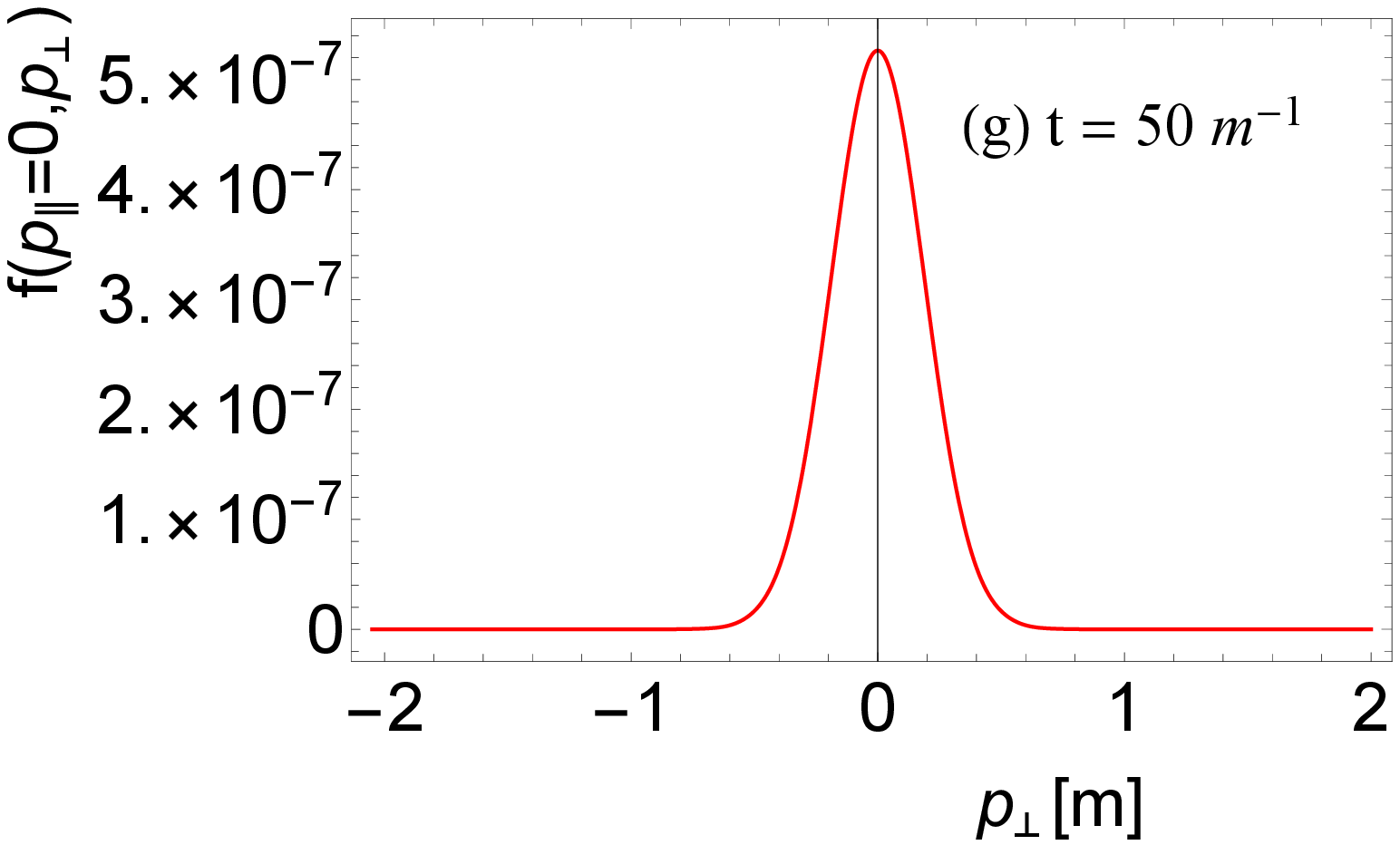}
\includegraphics[width = 2.55in]{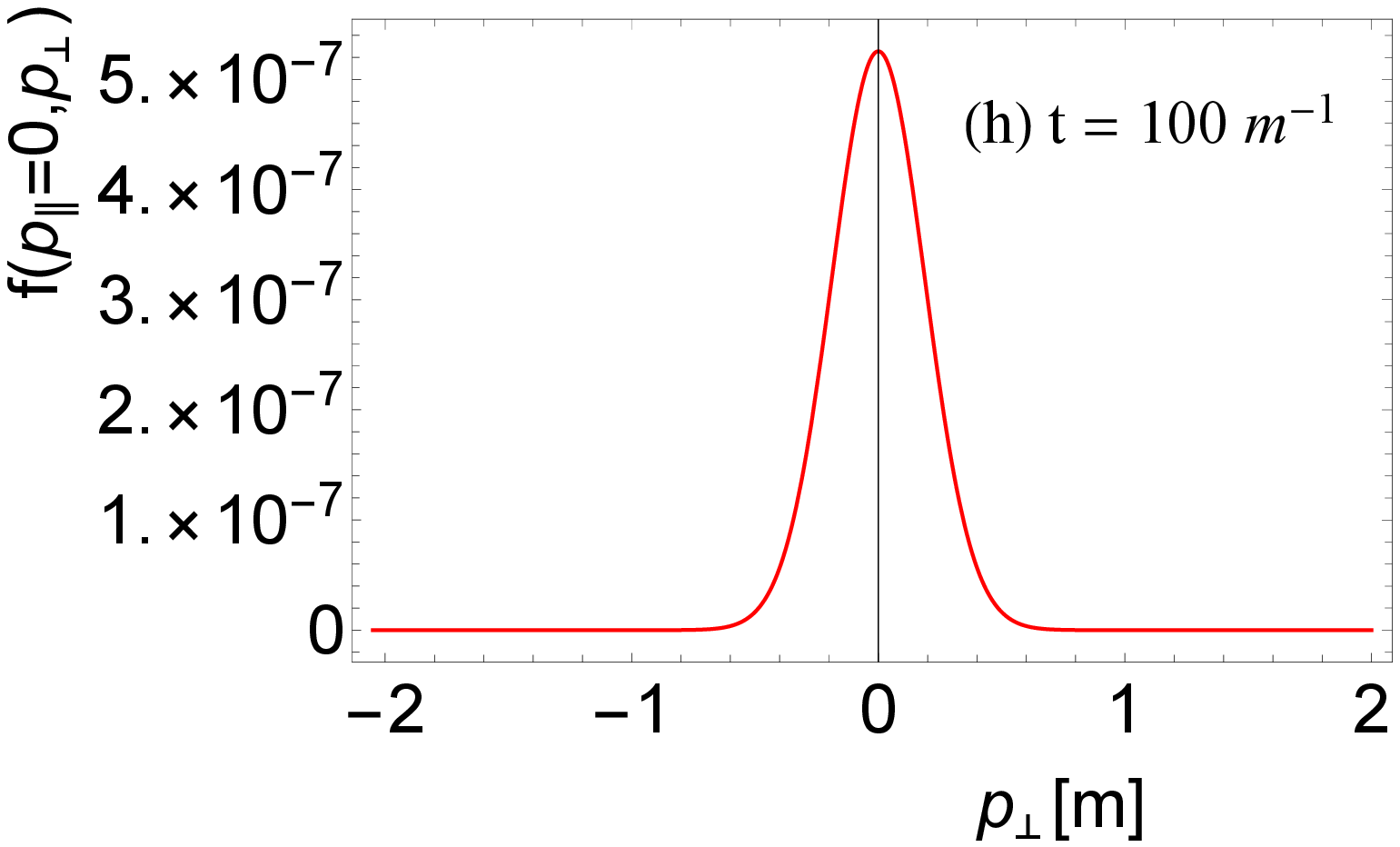}
}
\caption{Time evolution of quasi-particle distribution function in   the transverse  momentum space $f(p_\perp, t)$  at different times. The longitudinal momentum is considered to be zero, and all the units are taken in electron mass units.The field parameters are  $E_0=0.2$ and $ \tau =10.$}
   	\label{FG:2}
\end{center}
\end{figure}
\subsection{Transverse momentum spectrum }
Time evolution of the transverse momentum spectrum of created quasi-particle pairs are depicted in the figure \ref{FG:2}.The shape of the transverse
momentum distribution is changed when the electric field is switched on. As we see in the figure \ref{FG:2}(c),(d) at the transient stage the peak at $p_\perp = 0$ splits into weakly
pronounced peaks at $p_\perp \approx  \pm 0.67m$ and $p_\perp \approx  \pm 0.84m.$
At the asymptotic time (REPP stage \cite{46}), that splitting disappears and distribution function $f(p_\perp)$ shows the smooth structure with a peak at zero transverse momentum as shown in figure \ref{FG:2}(h).
\section{Conclusion}
We have analyzed the electron-positron pair creation from  the vacuum by a time-like Sauter pulse electric field. We have seen that the particle Longitudinal momentum distributions exhibit the oscillating structure at  a finite time where the electric field is nearly zero and this oscillating structure can be understood in the Dynamical Tunneling picture\cite{45}.
Dynamical tunneling can be understood as the process that develops in time inter-band dynamics in momentum representation for particles. There are different possibilities of particle tunneling through different channels in momentum space( or momentum states).
As time progresses, different possibilities can occur: (a) particle can tunnel at time $t$ with  $p_\parallel'-$value of momentum.
(b)At early stage,  particle with a lower momentum value ($p_\parallel < p_\parallel'$) can tunnel and then be accelerated to momentum $p_\parallel'$,
(c)Particle with a higher momentum value ($p_\parallel > p_\parallel'$) can tunnel at  time, $t_1$, and is then decelerated to get that momentum $p_\parallel'$ value.
Finally, when  the individual probability amplitudes of this process are added together, they result in quantum interference at time  $t$. 
At the asymptotic time, those processes do not share the same phase information and become random due to averaging over particle paths that do not show quantum interference effects, which leads to a smooth distribution function.
However, multiphoton absorption is not seen for the value of the Keldysh parameter $\gamma < 1,$ but it is believed to give a smaller contribution in comparison to what we get from tunneling. Since, in our calculation, it is not much less than one ($\gamma = 0.5$). As a result, another process of particle creation with momentum $p_\parallel$, namely the multiphoton process, cannot be ruled out. These are possibilities, and extensive quantitative  analyses of   these processes  are in progress\cite{DEEPAK:2023}.
The transverse momentum distribution shows only the splitting of smooth structure and the absence of quantum interference, which is an obvious interference effect that occurs only in the direction of the electric field.
\newline
A more detailed investigation of the  momentum distribution of created particles at the finite time is currently underway \cite{DEEPAK:2023}.

\section{Acknowledgments}
DEEPAK gratefully acknowledge the financial support from Homi Bhabha National Institute (HBNI) for carrying out this research work. Deepak also  thanks the organizers of the QED Laser Plasmas International Workshop 2022, Dresden for the possibility to present our work.

\end{document}